\documentclass[%
11pt,
preprint,
 nofootinbib,
 notitlepage,
 amsmath,amssymb,
 aps,
prd,
longbibliography
]{revtex4-2}

\usepackage{graphicx}% Include figure files
\usepackage{dcolumn}% Align table columns on decimal point
\usepackage{bm}% bold math
\usepackage[bookmarks=true,colorlinks=true,linkcolor=blue,unicode=true]{hyperref}% add hypertext capabilities
\usepackage[mathlines]{lineno}% Enable numbering of text and display math
\usepackage{xcolor} % for textcolor
\usepackage{color} % for color{red}

\newcounter{YJC}

\begin{document}

%\preprint{APS/123-QED}

\title{Split of the pseudo-critical temperatures of chiral and confine/deconfine transitions by temperature gradient}% Force line breaks with \\

\author{Ji-Chong Yang}
\email{yangjichong@lnnu.edu.cn}
\thanks{Corresponding author}
\author{Wen-Wen Li}
\email{lww1514148@163.com}
\author{Chong-Xing Yue}
\email{cxyue@lnnu.edu.cn}

\affiliation{Department of Physics, Liaoning Normal University, Dalian 116029, China}
\affiliation{Center for Theoretical and Experimental High Energy Physics, Liaoning Normal University, Dalian 116029, China}

\date{\today}% It is always \today, today,

\begin{abstract}
Searching of the critical endpoint~(CEP) of the phase transition of Quantum Chromodynamics~(QCD) matter in experiments is of great interest. 
The temperature in the fireball at a collider is location dependent, however, most theoretical studies address the scenario of uniform temperature. 
In this work, the effect of temperature gradients is investigated using lattice QCD approach. 
We find that the temperature gradient catalyzes chiral symmetry breaking, meanwhile the temperature gradient increases the Polyakov loop in the confined phase but suppresses the Polyakov loop in the deconfined phase. 
Furthermore, the temperature gradient decreases the pseudo-critical temperature of chiral transition but increases the pseudo-critical temperature of the confine/deconfine transition.
It is also found that the temperature gradient can drive the system away from the singularity, implying that the CEP in the $T-\mu$ plane may move in the direction of the first-order phase transition in the phase diagram due to the temperature gradient.
\end{abstract}

\maketitle

%\tableofcontents

\section{\label{sec:1}Introduction}

The study of the phase diagram of Quantum Chromodynamics~(QCD) is motivated by the desire to understand the fundamental properties of strongly interacting matter under extreme conditions.
QCD is the theory that describes the strong interaction and the behavior of quarks and gluons, which are the building blocks of protons, neutrons, and other hadrons. 
By exploring the phase diagram, which represents the different phases of matter as a function of temperature and baryon chemical potential, one can gain insight into the behavior of matter at high temperatures and densities~\cite{Fukushima:2013rx,Busza:2018rrf,Bzdak:2019pkr}, such as those present in the early universe~\cite{Braun-Munzinger:2008szb,Braun-Munzinger:2015hba}, neutron stars~\cite{Page:2006ud,Lattimer:2015nhk}, and heavy-ion collisions~\cite{STAR:2010vob,Gupta:2011wh,Andronic:2017pug,Luo:2020pef,STAR:2021iop}. 
Not only that, the study of QCD phase transitions is one of the important ways to explore the color confinement.

The search of the QCD critical endpoint~(CEP) at a finite chemical potential and high temperature is one of the main tasks in relativistic heavy-ion experiments~\cite{Stephans:2006tg,NA49-future:2006qne,Gazdzicki:2008kk,STAR:2010vob,Luo:2017faz}, and has attracted the attention of theoretical studies~\cite{Stephanov:2004wx,Fodor:2004nz,Gavai:2004sd,Antoniou:2006zb,Asakawa:2008ti,Stephanov:2011zz,Isserstedt:2019pgx,Fu:2019hdw,Gao:2020fbl,Bernhardt:2021iql}.
The precise determination of phase transition temperatures and the location of the CEP in QCD remains challenging. 
Experimental observations, provide indirect information about the QCD phase diagram through the analysis of particle spectra, yields, and correlations. 
These experimental observables can be used to infer the properties of the system, however, a direct mapping between experimental observables and theoretical phase transition temperatures is not straightforward and requires careful analysis and comparison.
There are other obstacles in the middle of theory and experiment, for example, phase transition theory studies mainly focus on equilibrium states, while there are non-equilibrium effects in experiments, and also the fact that the experimental situation is much more complicated, with issues such as finite volumes, position-dependent temperatures and baryon number densities~\cite{STAR:2021iop,Zheng:2021pia}.

\begin{figure}[htbp]
\begin{center}
\includegraphics[width=0.6\hsize]{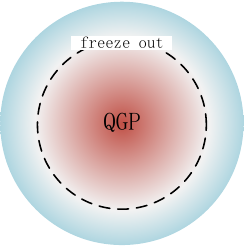}
\caption{\label{fig:fireball}A sketch of the temperature distribution of a fireball in a heavy ion collision.
The temperature decreases from inner to outer.
The dashed line is the chemical freeze out surface which overlaps with the isothermal surface of equilibrium phase transition in the case of uniform temperature~\cite{Zheng:2021pia}.}
\end{center}
\end{figure}
As shown in Fig.~\ref{fig:fireball}, at a heavy ion collider, the temperature of an instantaneous fireball decreases from inner to outer.
Typically, at a heavy ion collider, the quark-gluon-plasma~(QGP) and the hadronic phase are separated by a dynamical phase transition surface in a fireball, which is located in a region of non-uniform temperature.
It has been studied using an Ising-like effective potential that the temperature gradient will lift the critical temperature~\cite{Zheng:2021pia}.
Ref.~\cite{Zheng:2021pia} studies the situation of a brick cell of the fireball, in which there is a difference in the spatial distribution of temperature, and the stationary phase transition surface
in a steady temperature non-uniform system is studied.
In this work, a similar situation is considered using the lattice QCD approach with $N_f=2$~(two degenerate flavors) dynamic staggered fermions.
The non-uniform temperature is introduced with the help of non-uniform lattice spacings.

In hot QCD, there are two types of phase transitions in the $T-\mu$ plane where $T$ is the temperature and $\mu$ denotes the chemical potential. 
In the case of chiral fermions, there is the chiral symmetry phase transition, and the corresponding order parameter is the chiral condensation. 
In the case where dynamic fermions are absent, there is the confine/deconfine phase transition, the corresponding order parameter is the Polyakov loop~\cite{Polyakov:1978vu}. 
For finite fermion mass, both transitions are crossovers. 
For a long time, these two transitions were considered to occur simultaneously, except for the case of quarkyonic phase in the high baryon density region~\cite{McLerran:2007qj}. 
The pseudo-critical temperatures of the two transitions are close to each other in the lattice studies~\cite{Kogut:1982rt,Fukugita:1986rr,Karsch:1994hm,Digal:2000ar,Digal:2002wn,Bernard:2004je,Cheng:2006qk,Aoki:2006we,Aoki:2006br,Borsanyi:2010bp}, and the discrepancy between the two has been attributed to the nature of crossovers.
Studies have been devoted to investigating the interplay of the two in order to understand the nature of the phase transition of hot QCD matter~\cite{Polyakov:1978vu,tHooft:1977nqb,Casher:1979vw,Banks:1979yr,Braun:2009gm,Xu:2011wn,Xu:2012hq,Miura:2010xu}.
In this work, the chiral condensation, the Polyakov loop as well as the pseudo-critical temperature are studied with the focus on the shift of the pseudo-critical temperature.
However, we find that, in the case of zero baryon density, the temperature gradient causes the pseudo-critical temperatures of the two transitions to move away from each other, which is a phenomenon worth noting.

The remainder of the paper is organized as follows, the model and the parameters are introduced in section~\ref{sec:2}, the numerical results of lattice simulation with temperature gradient are shown in section~\ref{sec:3}, section~\ref{sec:4} summarizes our main conclusions.

\section{\label{sec:2}Lattice setup and parameters}

\begin{figure}[htbp]
\begin{center}
\includegraphics[width=0.48\hsize]{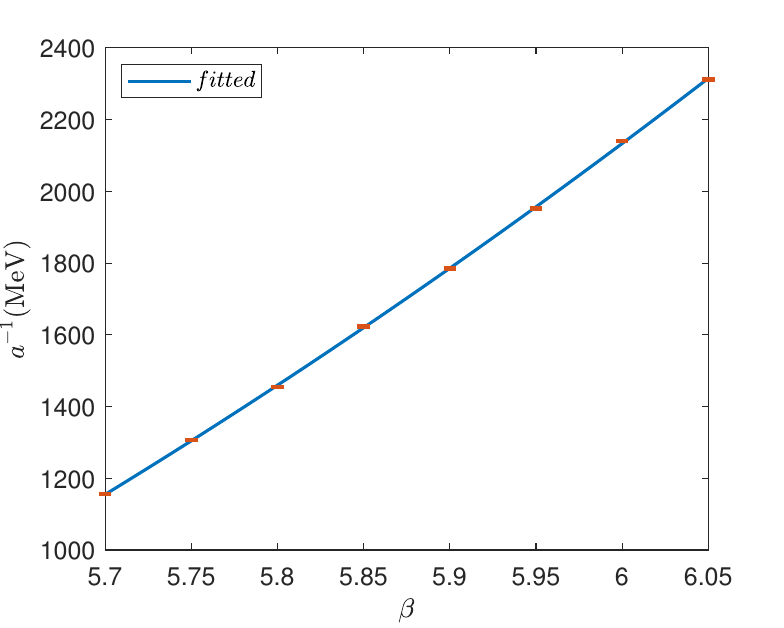}
\includegraphics[width=0.48\hsize]{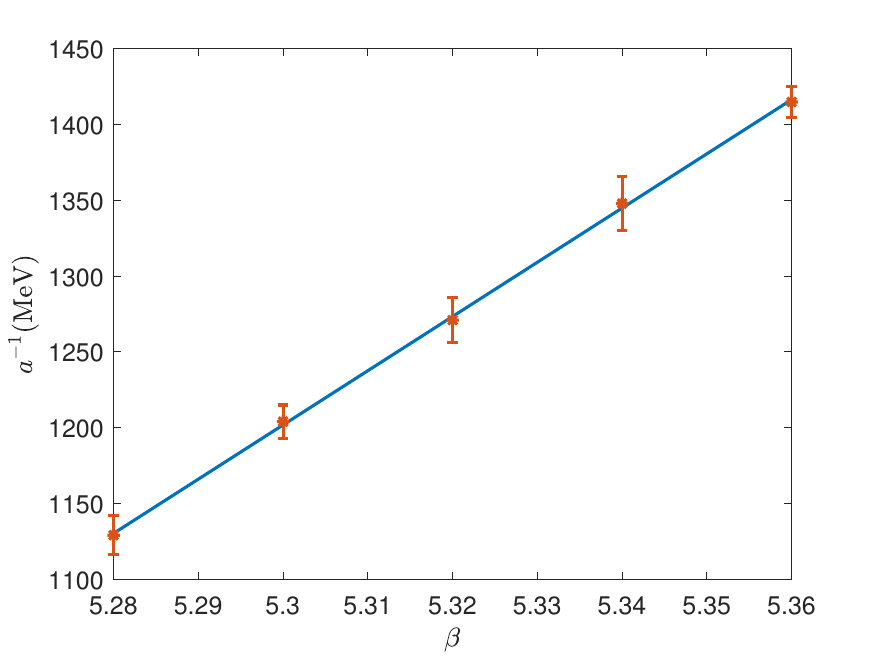}
\caption{\label{fig:latticespacing}The lattice spacing as a function of $\beta$, and the fitted results. The left panel is for the quenched approximation, the right panel is for the case when dynamic fermions are turned on.}
\end{center}
\end{figure}

In this work, both the case of the quenched approximation and the case of $N_f=2$ dynamic fermions are studied.
When the dynamic fermions are turned on, the Kogut-Susskind staggered fermions~\cite{Kogut:1974ag,Kluberg-Stern:1983lmr,Morel:1984di} are used, the action is $S=S_G+S_q$, and,
\begin{equation}
\begin{split}
&S_G=\frac{\beta(n_z)}{N_c}\sum _n \sum _{\mu>\nu}{\rm Retr}\left[1-\bar{U}_{\mu\nu}(n)\right],\\
&S_q=\sum _n \left(\sum _{\mu}\sum _{\delta = \pm \mu}\bar{\chi}(n) U_{\delta}(n)\eta _{\delta}(n)\chi (n+\delta) \right.\\
&\left.+2am\bar{\chi}\chi \right),
\end{split}
\label{eq.2.1}
\end{equation}
where $a$ is the lattice spacing, $m$ is the fermion mass with $am=0.1$ which is heavier than the physical mass, but lighter as a computational resource needed for exploratory research, $U_{\mu}=e^{iaA_{\mu}}$, $\eta _{\mu}(n)=(-1)^{\sum _{\nu<\mu}n_{\nu}}$, $U_{-\mu}(n)=U_{\mu}^{\dagger}(n-\mu)$, $\eta _{-\mu}=-\eta _{\mu}(n-\mu)$, and $\bar{U}_{\mu\nu}(n)=\left(U_{\mu,\nu}(n)+U_{\mu,-\nu}(n)+U_{-\mu,\nu}(n)+U_{-\mu,-\nu}(n)\right)/4$ where $U_{\mu,\nu}(n)=U_{\mu}(n)U_{\nu}(n+\mu)U_{-\mu}(n+\nu)U_{-\nu}(n)$.
The rational hybrid Monte Carlo~\cite{Clark:2003na,Clark:2006wp} is used to implement the `forth root trick'.

Note that, $\beta$ is introduced as a function of $n_z$, so that the lattice spacing is a function of $n_z$.
The effect of simulating a temperature gradient along the $z$-axis is achieved by assigning different lattice spacings to different $z$-slices along the $z$-axis, i.e., the whole volume is a number of $z$-slices with different temperatures stacked together. 
It is worth noting that in this way, for different $z$-slices, the lattice spacings, the extents in $x-y$ directions, and quark masses are also different. 
In this work, we ignore these differences mentioned above, and assume that the main factor of the variation due to different lattice spacing is from the difference in temperature.

\begin{figure}[htbp]
\begin{center}
\includegraphics[width=0.5\hsize]{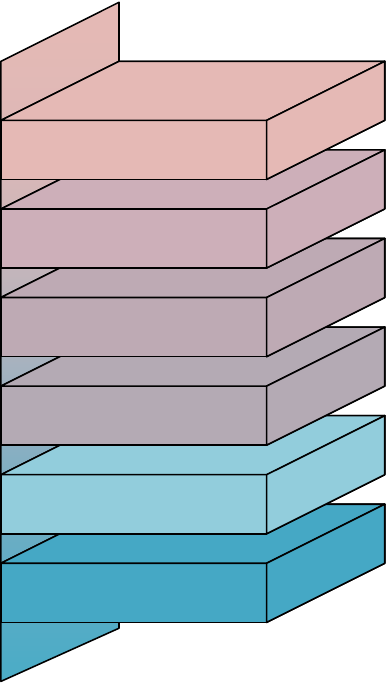}
\caption{\label{fig:setup}It is assumed that the different $z$-slices have already reached some temperature, for example, each $z$-slice is in contact with some external heat source, and then the $z$-slices are simply stacked on top of each other.}
\end{center}
\end{figure}
In this paper, we do not consider how gradients in temperature are formed, but only what effects temperature gradients have when they occur. 
That is, we are not giving two different temperatures to the ends of the $z$-axis so that the entire volume reaches thermal equilibrium. 
We assume that the different $z$-slices have reached some temperature (e.g., contact with some external heat source), and then simply stack the $z$-slices on top of each other, as shown in Fig.~\ref{fig:setup}.
In the Monte Carlo simulation, the entire lattice is considered as a whole. 
More specifically, we use the action defined in Eq.~(\ref{eq.2.1}), upon which the molecular dynamics equations are built. 
This action includes the gauge links connecting different z-slices. 
It differs from an ordinary one only in that the $\beta$ is location dependent.

To find out the temperatures corresponding to different values of $\beta$, matching is carried out using a $L_x\times L_y \times L_z \times L_{\tau} = 12^3\times 48$ lattice, where $L_{x,y,z,\tau}$ are extents at different directions.
For the quenched approximation, $\left(200+9800\right)$ molecular dynamics time units~(TUs) are simulated for each $\beta$ with $\beta =5.7 + 0.05 k$, where $0\leq k \leq 7$ are integers, $200$ trajectories of the $10000$ configurations are used for thermalization and $9800$ trajectories are measured.
When dynamic fermions are turned on, $\left(200+1000\right)$ configurations are generated for each $\beta$ with $\beta =5.28,5.3,5.32,5.34$ and $5.36$, where $200$ trajectories are used for thermalization and $1000$ trajectories are measured.
The lattice spacing $a$ is determined by measuring the static quark potential $V(r)$~\cite{Bali:1992ab,Bali:2000vr,Orth:2005kq} and then matching the `Sommer scale' $r_0$ to $0.5\;{\rm fm}$~\cite{Sommer:1993ce,Cheng:2007jq,MILC:2010hzw}.

\begin{table}
\begin{center}
\begin{tabular}{c|c|c|c|c|c}
\hline    
\multicolumn{3}{c}{Qunched approximation} & \multicolumn{3}{|c}{$N_f=2$ dynamic fermions}  \\
\hline    
$\beta$ & $r_0/a$ & $a^{-1}$ & $\beta$ & $r_0/a$ & $a^{-1}$ \\
& & (MeV) &  &  & (MeV) \\
\hline
$5.7$  & $2.932(3)$ & $1155(1)$ & $5.28$ & $2.8607(342)$ & $1129(13)$ \\
\hline
$5.75$ & $3.317(3)$ & $1307(1)$ & $5.3$  & $3.0503(278)$ & $1204(11)$ \\
\hline
$5.8$  & $3.693(3)$ & $1455(1)$ & $5.32$ & $3.2193(378)$ & $1271(15)$ \\
\hline
$5.85$ & $4.119(4)$ & $1623(2)$ & $5.34$ & $3.4155(449)$ & $1348(18)$ \\
\hline
$5.9$  & $4.532(5)$ & $1785(2)$ & $5.36$ & $3.5867(246)$ & $1415(10)$ \\
\hline
$5.95$ & $4.958(5)$ & $1953(2)$ & & & \\
\hline
$6.0$  & $5.435(6)$ & $2141(2)$ & & & \\
\hline
$6.05$ & $5.867(8)$ & $2312(3)$ & & & \\
\hline
\end{tabular}
\end{center}
\caption{\label{tab.matching}Results of matching for different values of $\beta$.}
\end{table}

The values of $\beta$ used in matching, and the corresponding $r_0$ and $a$ are shown in Table~\ref{tab.matching}.
In order to cover the values of $\beta$ where no matching was performed, and also to compute the derivative of the temperature w.r.t. $n_z$, the dependence of $a^{-1}$ on $\beta$ is fitted to a function, under the assumption that the range of $\beta$ is small enough and that the first few orders of the Taylor expansion are sufficient to characterize this relationship.
In the case of quenched approximation, a bilinear function is sufficient.
When dynamic fermions are turned on, the range of $\beta$ is small enough that it is sufficient to consider only the linear order of the Taylor expansion.
The results of the fittings are shown in Fig.~\ref{fig:latticespacing}, and can be written as,
\begin{equation}
\begin{split}
&\left.\frac{a^{-1}}{\rm MeV}\right|_{\rm Quench}=19451.4-9357.7\beta + 1078.6 \beta^2,\\
&\left.\frac{a^{-1}}{\rm MeV}\right|_{N_f=2}=-17772.2 + 3580.0\beta.\\
\end{split}
\label{eq.afitres}
\end{equation}

We focus on the Polyakov loop and chiral condensation and their susceptibilities. 
The quantities are defined as~\cite{Bazavov:2011nk},
\begin{equation}
\begin{split}
&P=\frac{1}{N_cL_xL_yL_z}\sum _{\vec{n}} \prod _{n_{\tau}}U_{\tau}(\vec{n}),\\
&\chi _P = a^3L_xL_yL_z\left(\langle P ^2 \rangle - \langle P\rangle ^2\right),\\
&c=\frac{1}{a^3L_xL_yL_zL_{\tau}}{\rm tr}[D^{-1}],\\
&\chi _{disc}=\frac{1}{a^2L_xL_yL_zL_{\tau}}\left(\langle {\rm tr}\left[D^{-1}\right]^2\rangle-\langle {\rm tr}\left[D^{-1}\right]\rangle^2\right),\\
\end{split}
\label{eq.cqdefine}
\end{equation}
where $P$ is the bare Polyakov loop, $c$ is the bare chiral condensation, $\chi _P$ is the susceptibility of $P$, and $\chi _{disc}$ is the disconnected susceptibility of the bare chiral condensation.
In the case of the quenched approximation, we use $\chi _{|P|} =  a^3L_xL_yL_z \left(\langle\left| P\right| ^2 \rangle - \langle \left|P\right|\rangle ^2\right)$.
When there is a temperature gradient, we focus on simulations on a lattice with $L_{x}L_yL_zL_{\tau} = 12^3\times 6$. 
In the uniform temperature case, $2000+(100+49900)\times 8$ TUs are simulated, where $50000$ TUs are simulated for each value of $\beta$ sequentially with $\beta$ growing, except for the first value of $\beta$ that $52000$ TUs are simulated. 
The last $49900$ configurations for each value of $\beta$ are measured. 
In the quenched approximation, $\beta = 5.7 + 0.05\times k$, in the $N_f=2$ case, $\beta=5.29 + 0.01\times k$, both with $0\leq k \leq 7$.
\begin{figure}[htbp]
\begin{center}
\includegraphics[width=0.48\hsize]{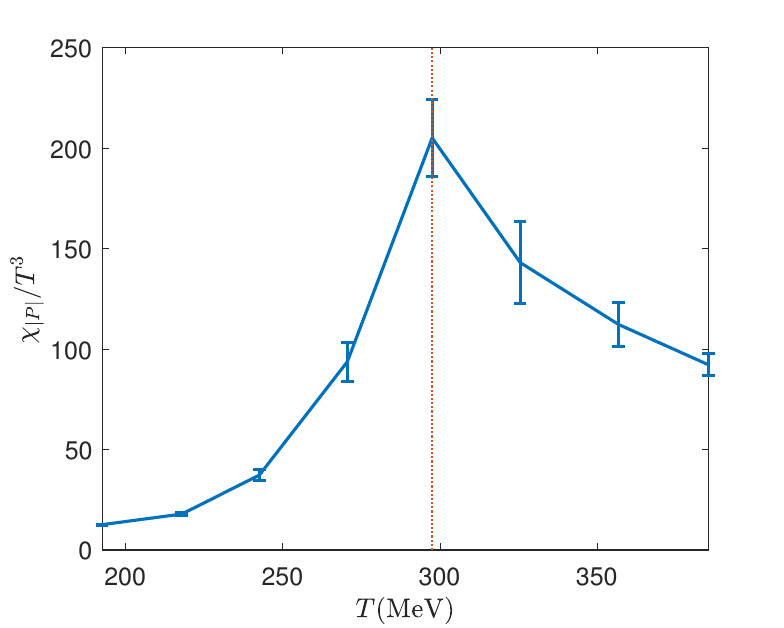}
\includegraphics[width=0.48\hsize]{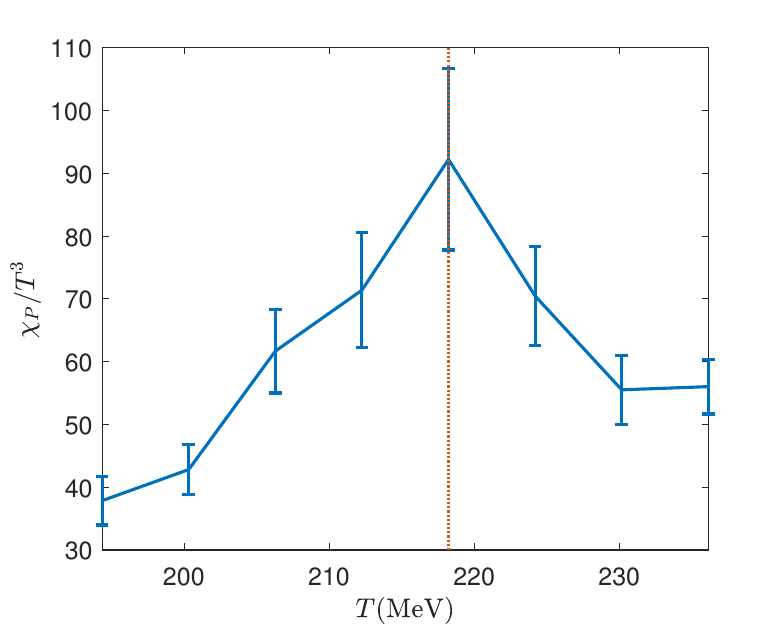}
\includegraphics[width=0.48\hsize]{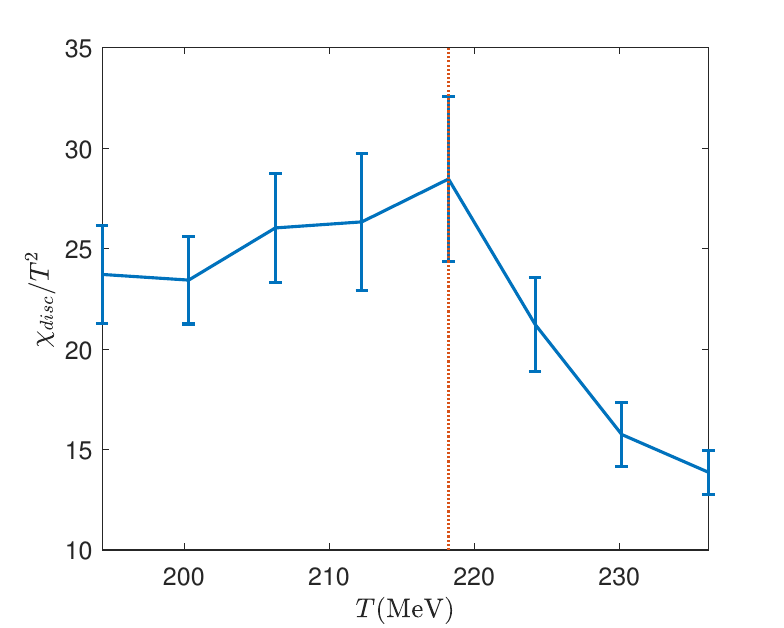}
\caption{\label{fig:suspuniform}The susceptibilities as functions of temperature. 
The top left panel is $\chi _{|P|}$ in the quenched approximation. 
The top right panel is $\chi _P$ and the bottom panel is $\chi _{disc}$ when dynamic fermions are turned on.
The dotted line in the top left panel shows the critical temperature of the confined/deconfine phase transition, and in the other two panels show the pseudo-critical temperatures of the transitions.}
\end{center}
\end{figure}
At the above parameters, the susceptibilities $\chi _P$ and $\chi _{disc}$ are shown in Fig.~\ref{fig:suspuniform}. 
The (pseudo-)critical temperatures of the transitions can be obtained as locations of the peaks of the susceptibilities.

Throughout this paper, we use $\sigma = \sqrt{2 \tau _{\rm int}} \sigma _{\rm jk}$ as statistical error~\cite{Gattringer:2010zz}, where $\tau _{\rm int}$ is the separation of TUs such that the two configurations can be regarded as independent, and $\sigma _{\rm jk}$ is statistical error calculated using `jackknife' method. $\tau _{\rm int}$ is calculated by using `autocorrelation' with $S=1.5$~\cite{Wolff:2003sm} on the bare Polyakov loop or the bare chiral condensation depending on whether we are measuring quantities w.r.t. Polyakov loop or chiral condensation.
The one exception is the lattice spacing, where the statistical error is calculated using $\tau _{\rm int}$ estimated from the bare Polyakov loop.

\section{\label{sec:3}Numerical results for the case of temperature gradient}

Since in our simulation, $\beta$ is a function of $z$, the quantities in a $z$-slice are also interesting.
They are defined as,
\begin{equation}
\begin{split}
&P(n_z)=\frac{1}{N_cL_xL_y}\sum _{n_{x,y}} \prod _{n_{\tau}}U_{\tau}(\vec{n}),\\
&c(n_z)=\frac{1}{a^3L_xL_yL_{\tau}} \sum _{n_{x,y,\tau}}D^{-1}(n|n).\\
\end{split}
\label{eq.cqzdefine}
\end{equation}

\begin{figure}[htbp]
\begin{center}
\includegraphics[width=0.6\hsize]{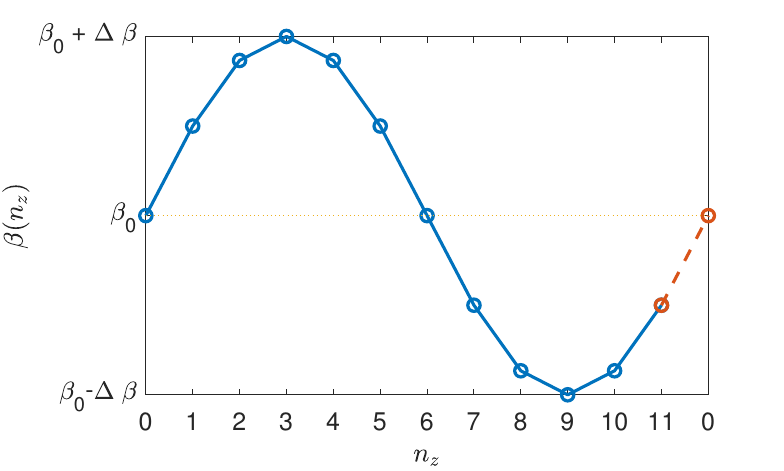}
\caption{\label{fig:betaz}A sketch of $\beta(n_z)$ function.
A periodic function is used to avoid possible effects from the boundary.
Note that, at $n_z=0$ and $6$, $\beta(n_z)=\beta_0$, and the absolute values of the temperature gradients are the same.}
\end{center}
\end{figure}
The origin of the coordinate is chosen such that $0\leq n_z \leq 11$, and,
\begin{equation}
\begin{split}
&\beta(n_z)=\beta_0 + \Delta \beta \sin (\frac{2n_z\pi }{L_z} ),\\
\end{split}
\label{eq.betacase1}
\end{equation}
where $L_z=12$ is the extent on the $z$ direction.
A sketch of $\beta(n_z)$ is shown in Fig.~\ref{fig:betaz}
The reason we use a periodic function for $\beta(n_z)$ is to avoid the analysis of possible effects from the boundary.
In the case of quenched approximation, the fermion mass is infinite, and thus receives no effect from the change of lattice spacing. 
As a way to explore the effect of the temperature gradient, we intentionally choose a large temperature gradient, only $\Delta \beta =0.5$ is studied.
In the case of $N_f=2$ dynamic fermions, to study the shift of the pseudo-critical temperature, four cases are considered such that $\Delta \beta=0.05, 0.1, 0.15$ and $0.2$.
The values of $\beta_0$ are as same as those of $\beta$ used in previous section, and the number of configurations for each $\Delta \beta$ are also as same as those in the previous section, therefore the results of the previous section consist as the case of $\Delta \beta = 0$.
The temperature gradient can be calculated as,
\begin{equation}
\begin{split}
&\frac{\partial T}{\partial z} = \frac{2\Delta \beta \pi }{aL_{\tau}L_z} \cos (\frac{2n_z\pi}{L_z})\frac{\partial a^{-1}}{\partial \beta}.\\
\end{split}
\label{eq.temperaturegradient}
\end{equation}

\subsection{\label{sec:3.0}Distribution of Polyakov loop and chiral condensation}

\begin{figure}[htbp]
\begin{center}
\includegraphics[width=0.48\hsize]{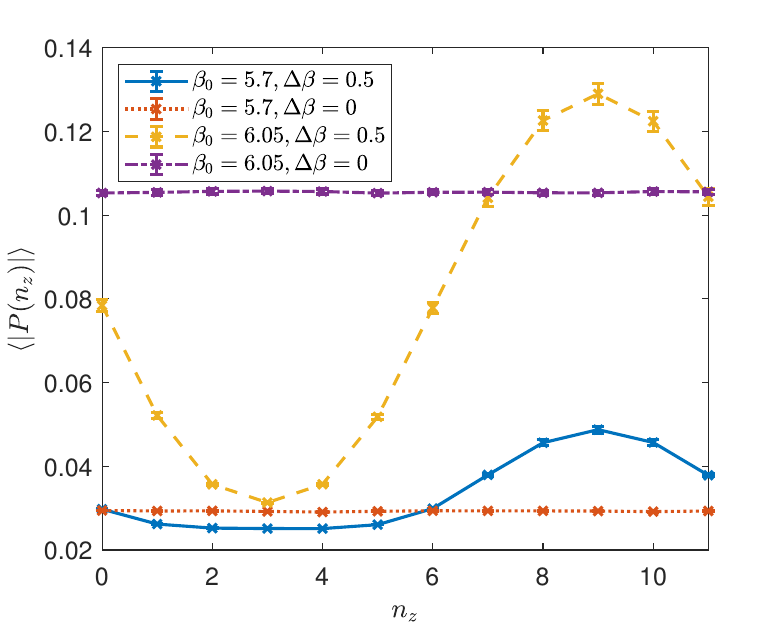}
\includegraphics[width=0.48\hsize]{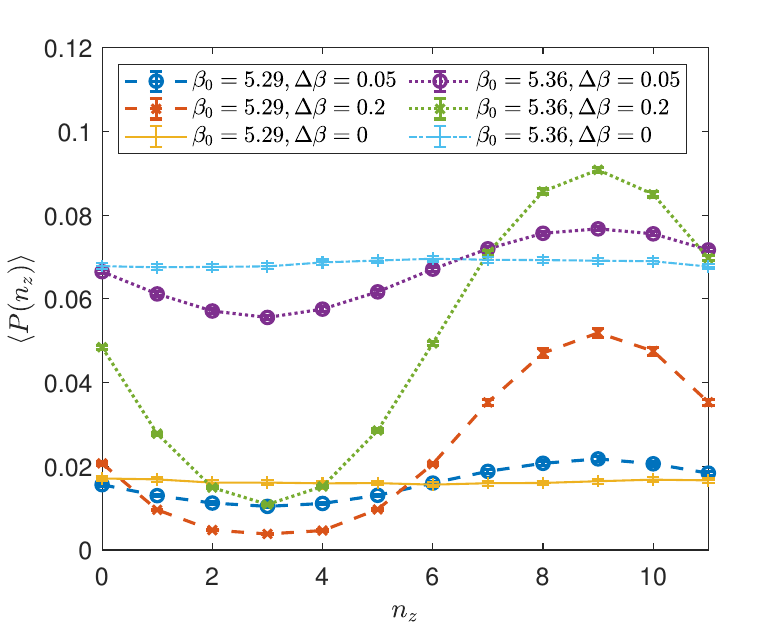}\\
\includegraphics[width=0.48\hsize]{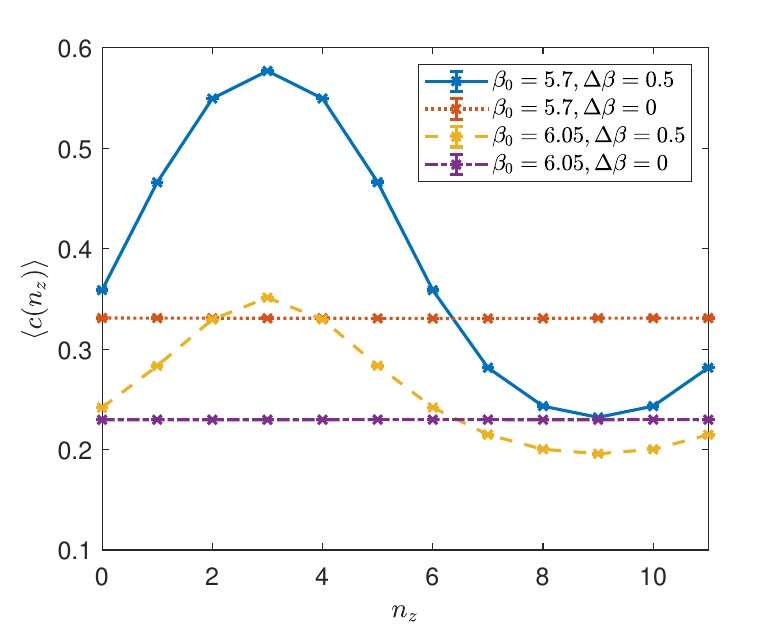}
\includegraphics[width=0.48\hsize]{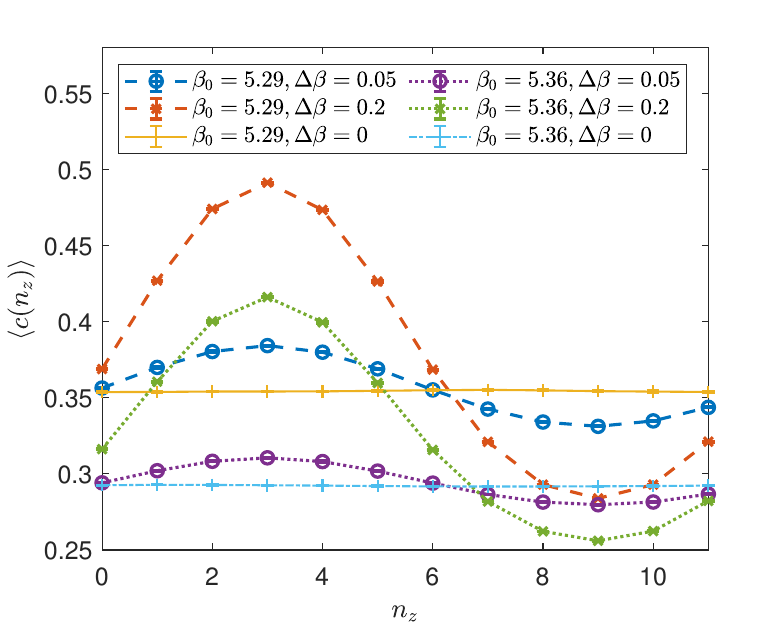}\\
\caption{\label{fig:zdistribution}$\langle c(n_z)\rangle$, $\langle |P(n_z)|\rangle$ and $\langle P(n_z)\rangle$ at different $n_z$, $\beta$ and $\Delta \beta$. 
The first row is for $P(n_z)$~($|P(n_z)|$ in the case of the quenched approximation), and the second row is for $c(n_z)$. 
The left panels are for the case of quenched approximation, and the right panels are for the case of $N_f=2$ dynamic fermions.}
\end{center}
\end{figure}

When $\beta$ is position dependent, both $P$ and $c$ become position dependent.
Examples of $P(n_z)$ and $c(n_z)$ at different $\beta$ and $\Delta \beta$ are shown in Fig.~\ref{fig:zdistribution}.
It can be seen that, the observables defined in a $z$-slice can generally reflect the temperature of the $z$-slice, i.e., for larger $\beta$, $P(n_z)$~(or $|P(n_z)|$) is larger, and $c(n_z)$ is smaller.

However, different $z$-slices when put together cannot be seen as simply stacked on top of each other as independent $z$-slices.
For example, at $n_z=0$, $\beta = \beta _0$ for different $\Delta \beta$.
It can be seen from Fig.~\ref{fig:zdistribution} that, there are differences in $P(n_z)$ and $c(n_z)$ even though they are in the $z$-slices with a same value of $\beta$.
That is, the distribution of temperature can affect nearby areas, which is exactly the effect of temperature gradients~(or higher order derivatives of the temperature distribution).

\begin{figure}[htbp]
\begin{center}
\includegraphics[width=0.48\hsize]{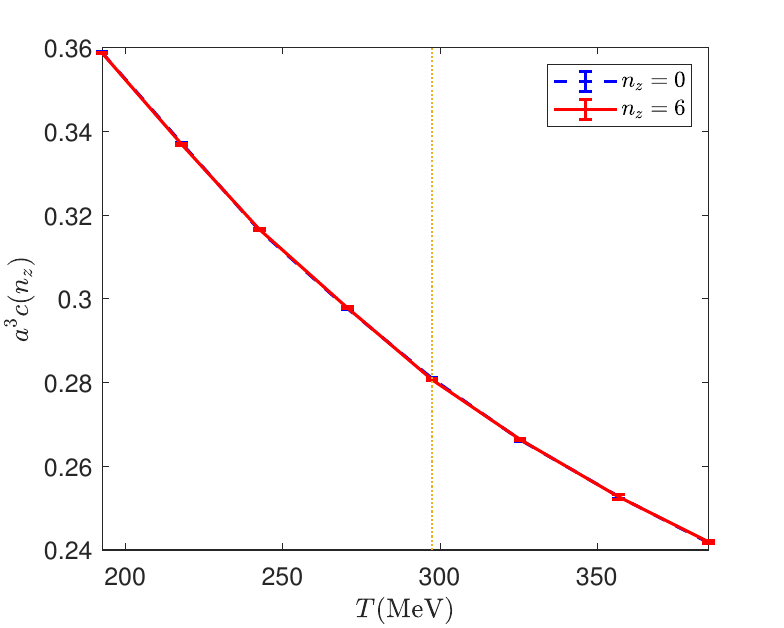}
\includegraphics[width=0.48\hsize]{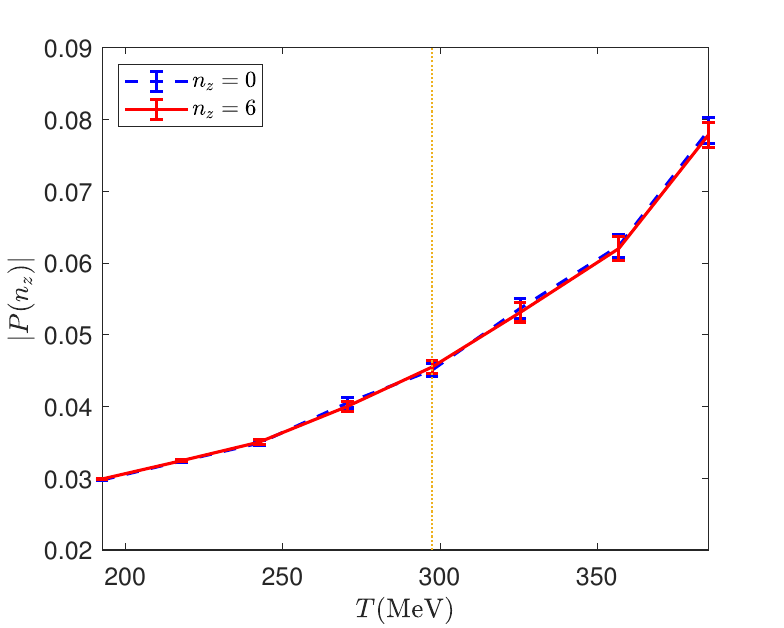}\\
\includegraphics[width=0.48\hsize]{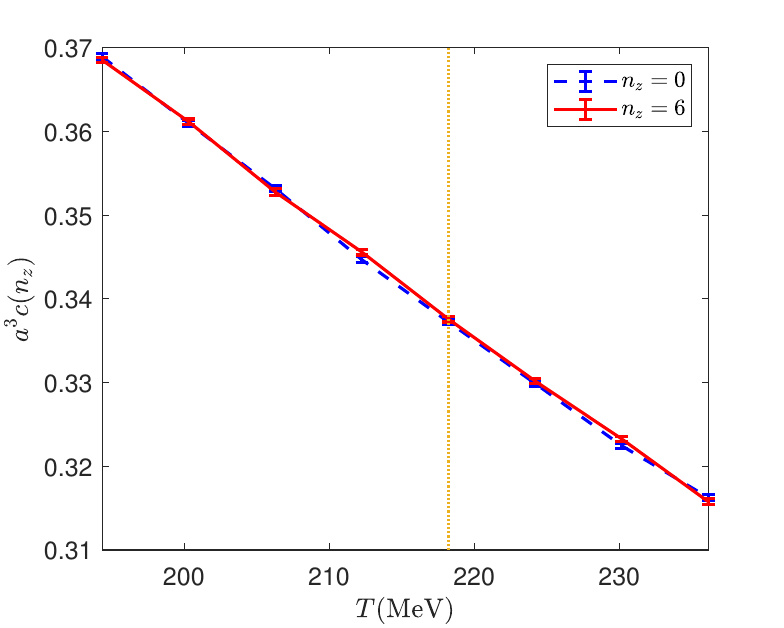}
\includegraphics[width=0.48\hsize]{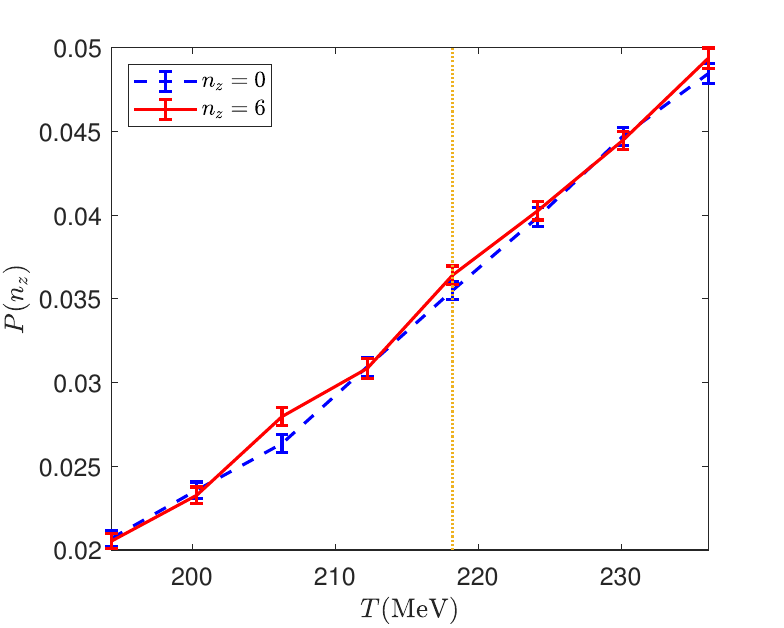}\\
\caption{\label{fig:06}$\langle c(n_z)\rangle$, $\langle |P(n_z)|\rangle$ and $\langle P(n_z)\rangle$ at $n_z=0$ and $6$. 
The first row is for the quenched approximation, where the left panel is $\langle c(n_z)\rangle$, the right panel is $\langle |P(n_z)|\rangle$. 
The second row is for the case when dynamic fermions are turned on, where the left panel is $\langle c(n_z)\rangle$, the right panel is $\langle P(n_z)\rangle$.
The dotted lines show the (pseudo)-critical temperatures of the transitions for the case of uniform temperature.
The {\bf x}-axes of the figures show the temperatures at $\beta_0$, i.e. temperatures at $n_z=0,6$.}
\end{center}
\end{figure}

Apart from that, for each pair of $z$-slices with $n_z=0,6$, $n_z=1,5$, $n_z=2,4$, $n_z=7,11$, and $n_z=8,10$, the $\beta$ and $|\partial \beta / \partial n_z|$ are the same for $z$-slices in the pairs.
It can be observed in Fig.~\ref{fig:zdistribution} that, $P(n_z)$ and $c(n_z)$ are generally the same for each pair of $z$-slices.
A naive conjecture is that neither the chiral condensation nor the Polyakov loop should depend on the direction of the temperature gradient. 
This can be verified by comparing the chiral condensation and the Polyakov loop at $n_z=0$ and $n_z=6$. 
For the quenched approximation, and for $N_f=2$ with $\Delta \beta=0.2$, the chiral condensation and the Polyakov loop are shown in Fig.~\ref{fig:06}.
It can be seen that the difference between $n_z=0$ and $n_z=6$ is negligible even when the temperature gradient is very large. 
Therefore, in the following, $n_z=0$ and $n_z=6$ are combined together, i.e., we use $P^{(\Delta \beta)}=\left(P(0)+P(6)\right)/2$, and $c^{(\Delta \beta)}=\left(c(0)+c(6)\right)/2$, in the case of the quenched approximation, we also considered $\tilde{P}^{(\Delta \beta)}=\left(\left|P(0)\right|+\left|P(6)\right|\right)/2$.

\begin{figure}[htbp]
\begin{center}
\includegraphics[width=0.48\hsize]{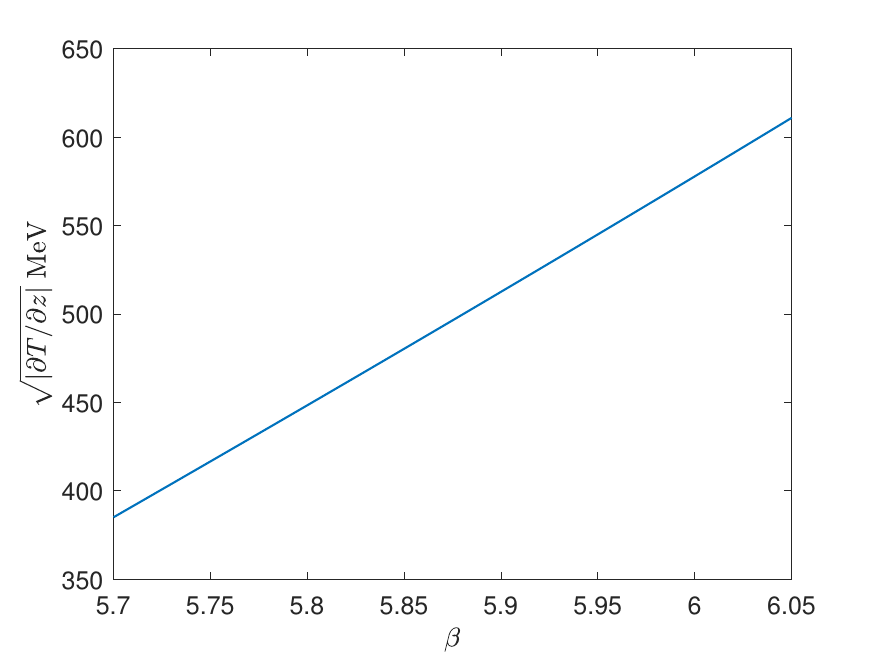}
\includegraphics[width=0.48\hsize]{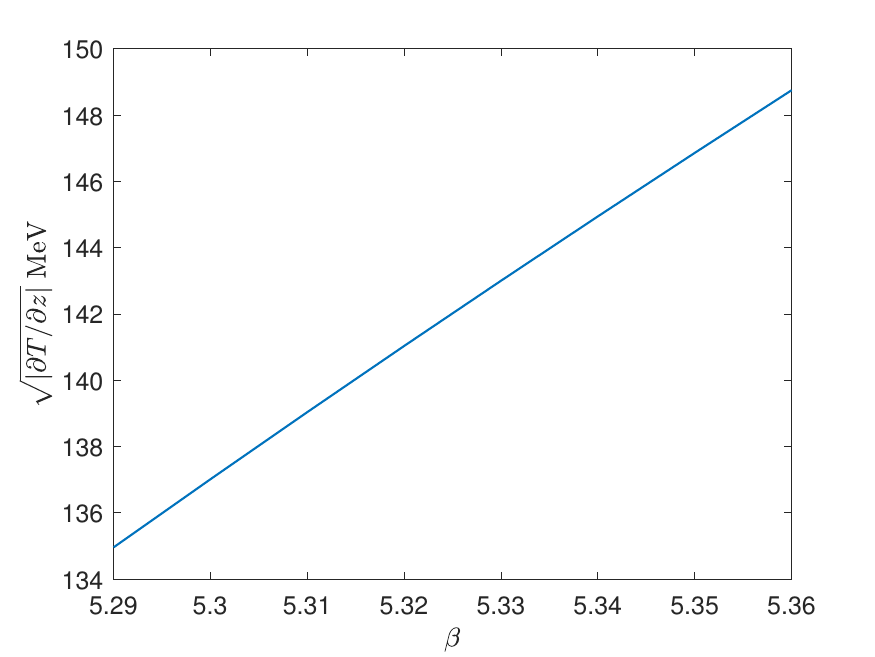}\\
\caption{\label{fig:tg}$\left|\partial T / \partial z\right|$ for the quenched approximation with $\Delta \beta=0.5$~(left panel) and for $N_f=2$ with $\Delta \beta = 0.05$~(right panel) at $n_z=0,6$ calculated using Eq.~(\ref{eq.temperaturegradientnz06}).
The cases for $N_f=2$ with $\Delta \beta = 0.1, 0.15$, and $0.2$ are two, three, and four times of the case for $\Delta \beta = 0.05$, respectively.}
\end{center}
\end{figure}
Note that, at $n_z=0,6$, $\beta(n_z)=\beta_0$, so $P^{(\Delta \beta)}$ and $c^{(\Delta \beta)}$ are Polyakov loops and chiral condensations at $\beta _0$ with different temperature gradients.
At $n_z=0,6$, 
\begin{equation}
\begin{split}
&\left|\frac{\partial T}{\partial z}\right| =\frac{\Delta \beta \pi}{36a} \left|\frac{\partial a^{-1}}{\partial \beta} \right|.
\end{split}
\label{eq.temperaturegradientnz06}
\end{equation}
In the quenched approximation, $\Delta \beta = 0.5$ corresponds to the range of the temperature gradient from $\left|\partial T/\partial z\right|=\left(385\;{\rm MeV}\right)^2$~($\beta=5.7$) to $\left|\partial T/\partial z\right|=\left(611\;{\rm MeV}\right)^2$~($\beta=6.05$).
In the case of $N_f=2$ and $\Delta \beta=0.05$, the range of temperature gradient is from $\left|\partial T/\partial z\right|=\left(135\;{\rm MeV}\right)^2$~($\beta=5.29$) to $\left|\partial T/\partial z\right|=\left(149\;{\rm MeV}\right)^2$~($\beta=5.36$).
The ranges of temperature gradient for the cases of $\Delta \beta = 0.1, 0.15$, and $0.2$ are two, three, and four times of the case for $\Delta \beta = 0.05$, respectively.
$\left|\partial T/\partial z\right|$ at $n_z=0,6$ with different $\beta$ are shown in Fig.~\ref{fig:tg}.

\subsection{\label{sec:3.1}Effect of temperature gradient on Polyakov loop and chiral condensation}

\begin{figure}[htbp]
\begin{center}
\includegraphics[width=0.48\hsize]{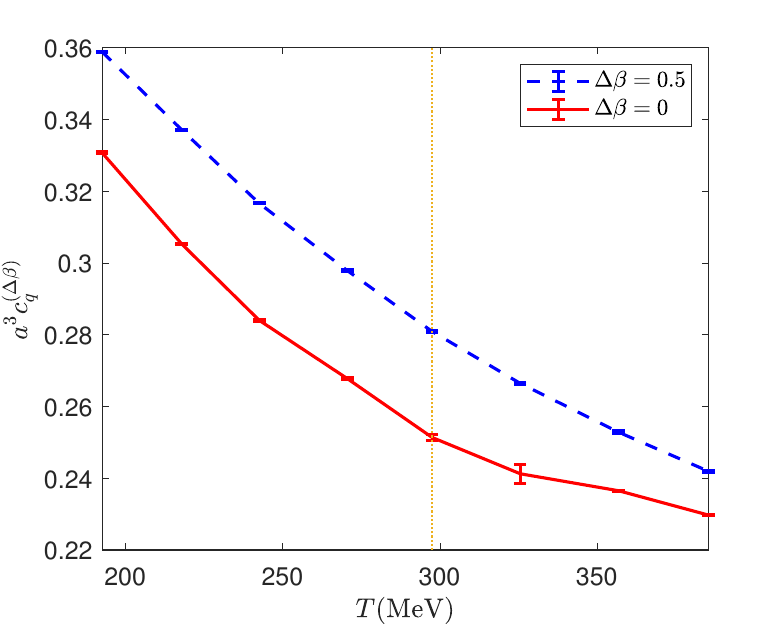}
\includegraphics[width=0.48\hsize]{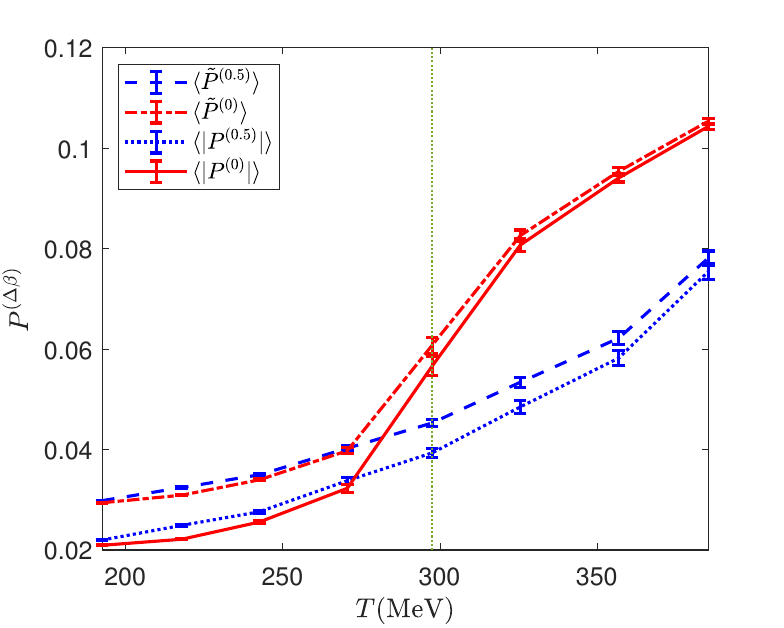}
\caption{\label{fig:dbq}$\langle c^{(\Delta \beta)} \rangle$~(left panel), $\langle |P^{(\Delta \beta)}|\rangle$ and $\langle \tilde{P}^{(\Delta \beta)}\rangle$~(right panel) as functions of $T$ in the quenched approximation. The dotted lines show the critical temperature of the confine/deconfine phase transition for the case of uniform temperature.}
\end{center}
\end{figure}
$\langle c^{(\Delta \beta)} \rangle$, $\langle |P^{(\Delta \beta)}|\rangle$ and $\langle \tilde{P}^{(\Delta \beta)}\rangle$ in the quenched approximation are shown in Fig.~\ref{fig:dbq}.
It can be shown that, the temperature gradient always increase the chiral condensation, while only suppressing the Polyakov loop in the deconfined phase.
Note that, in the quenched approximation, the temperature gradient is set to be large.
A noteworthy phenomenon is that, in the confined phase, a large temperature gradient barely affects the Polyakov loop.
Assuming that the distribution of temperature is continuous, the derivatives~(at every order) of temperature at a given location taken together essentially respond to the distribution of temperature in the region around that location. 
Therefore, the effect of the temperature gradient actually reflects the effect of temperature distribution at the subleading order.
When viewed as a whole, the phenomenon essentially comes from the distribution of temperature. 
Thus, the reason of the above phenomenon can be speculated to be that, in the confined phase and without dynamic fermions, the gluon field is hardly affected by the distant gluon field, while this is not true in the deconfined phase.

\begin{figure}[htbp]
\begin{center}
\includegraphics[width=0.48\hsize]{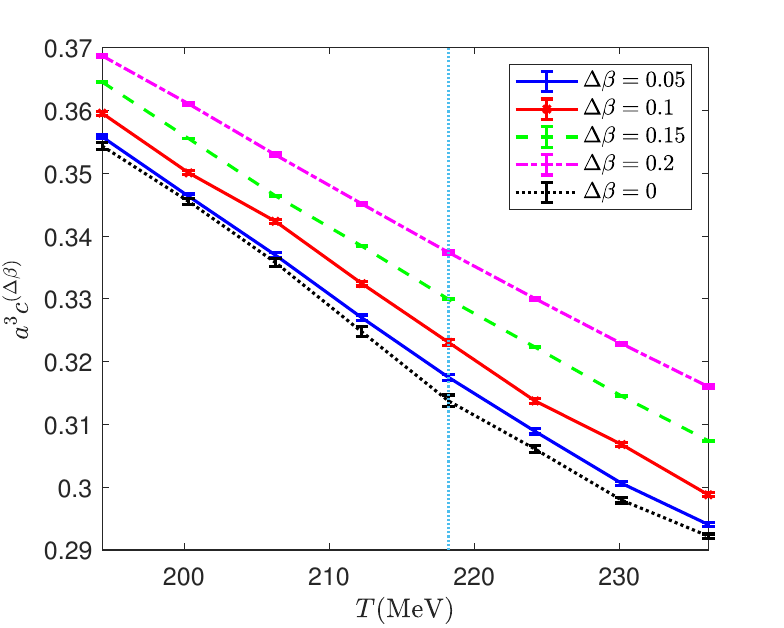}
\includegraphics[width=0.48\hsize]{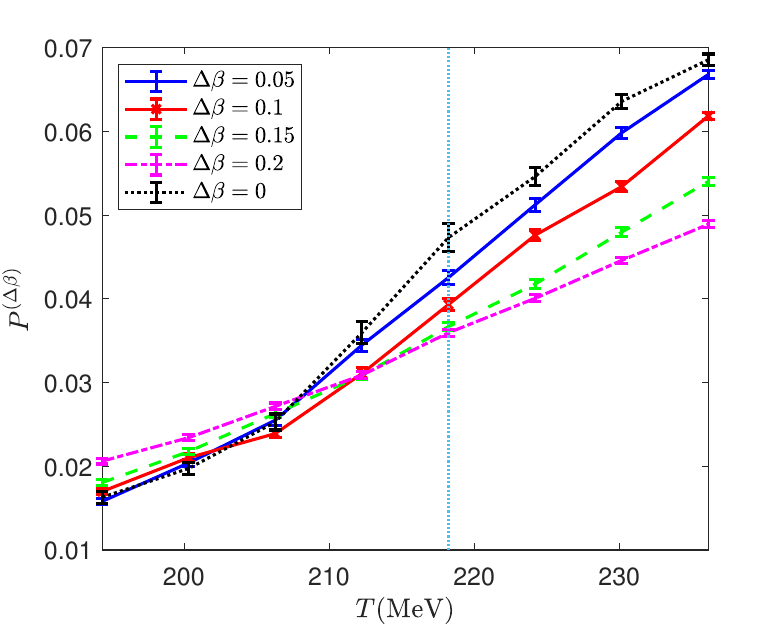}
\caption{\label{fig:polyakovandchiral}$\langle c^{(\Delta \beta)}\rangle$~(left panel) and $\langle P^{(\Delta \beta)} \rangle$~(right panel) at different temperatures and different temperature gradients.
The dotted lines show the pseudo-critical temperature of the transitions for the case of uniform temperature.}
\end{center}
\end{figure}

When dynamic fermions are turned on, the results are shown in Fig.~\ref{fig:polyakovandchiral}.
It can be seen from the results that for chiral condensation, a temperature gradient always leads to chiral symmetry breaking, and a larger temperature gradient leads to larger chiral symmetry breaking. 
Meanwhile, the temperature gradient can increase Polyakov loop at low temperatures and decrease Polyakov loop at high temperatures. 
Moreover, the curves of Polyakov loop versus temperature for different temperature gradients intersect with the curve in the uniform temperature case approximately near the pseudo-critical temperature of the confine/deconfine transition for the uniform temperature case.
Thus it can be concluded that the temperature gradient contributes to the deconfinement in the confined phase, and the opposite is true in the deconfined phase.
A noteworthy phenomenon is that in the confined phase, the temperature gradient can increase both chiral condensation and Polyakov loop.

\subsection{\label{sec:3.2}Shift of phase transition with \texorpdfstring{$N_f=2$}{Nf=2} staggered fermions}

To investigate the phase transition, the susceptibilities with the temperature gradient are defined as,
\begin{equation}
\begin{split}
&\chi^{(\Delta \beta)} _P = 2a^3L_x^2 \left(\langle \left(P^{(\Delta \beta)}\right) ^2 \rangle - \langle P^{(\Delta \beta)}\rangle ^2\right),\\
&\chi^{(\Delta \beta)} _{disc}=2a^4L_x^2L_{\tau}\left(\langle \left(c^{(\Delta \beta)}\right)^2\rangle-\langle c^{(\Delta \beta)}\rangle^2\right).\\
\end{split}
\label{eq.chidefine}
\end{equation}
In Eq.~(\ref{eq.chidefine}), the $n_z=0$ and $n_z=6$ $z$-slices are combined as a $12^2\times 2 \times 6$ lattice.
For the quenched approximation, $\chi^{(\Delta \beta)} _{|P|}$ and $\chi^{(\Delta \beta)} _{\tilde{P}}$ are defined as same as in Eq.~(\ref{eq.chidefine}) except that $P^{(\Delta \beta)}$ is substituted by $|P^{(\Delta \beta)}|$ and $\tilde{P}^{(\Delta \beta)}$.

\begin{figure}[htbp]
\begin{center}
\includegraphics[width=0.8\hsize]{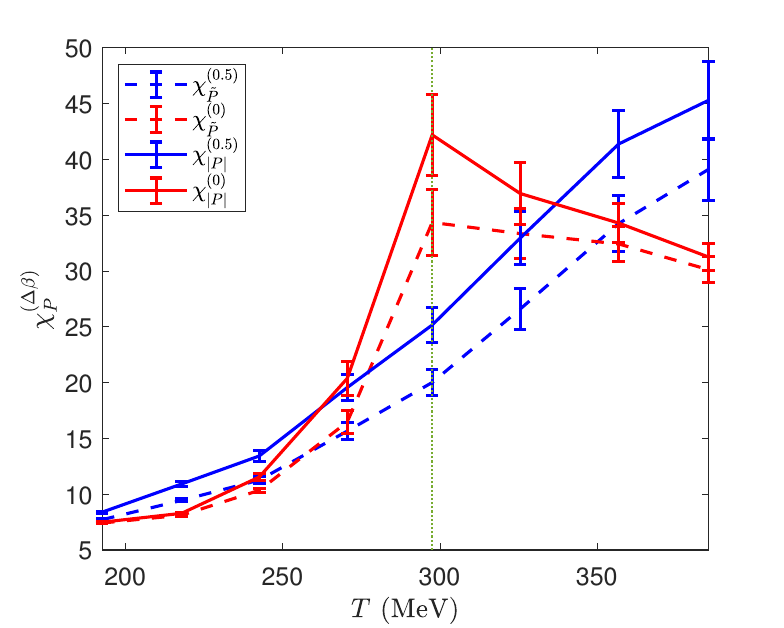}
\caption{\label{fig:quenchsusp}$\chi^{(\Delta \beta)} _{|P|}$ and $\chi^{(\Delta \beta)} _{\tilde{P}}$ as functions of $T$ in the quenched approximation.
The dotted lines show the critical temperature for the case of uniform temperature.}
\end{center}
\end{figure}

\begin{figure}[htbp]
\begin{center}
\includegraphics[width=0.8\hsize]{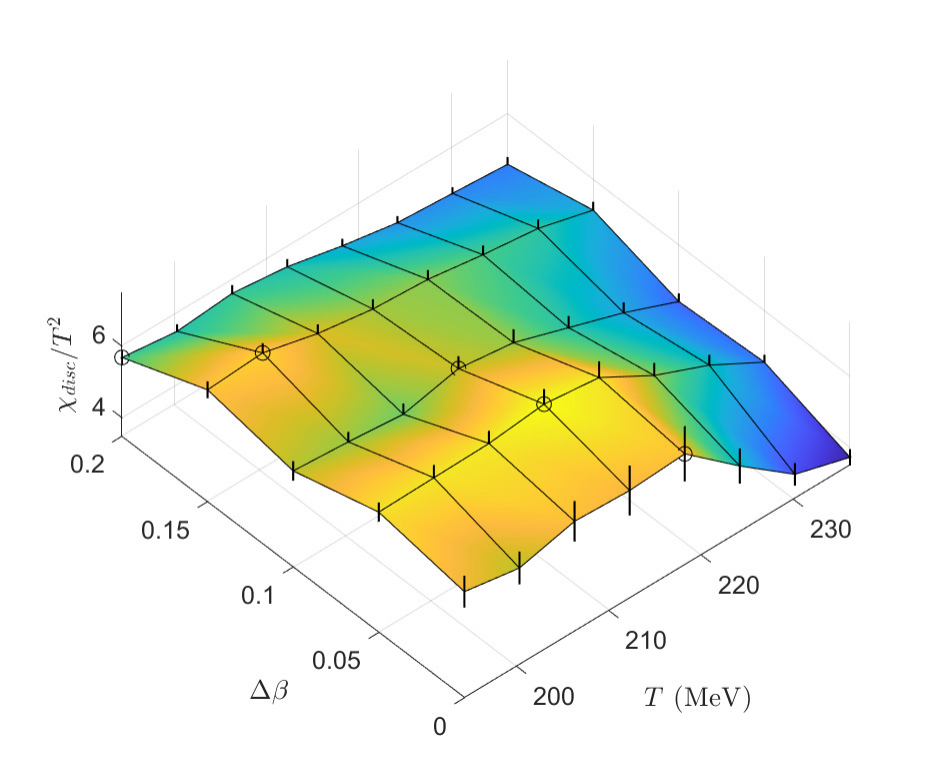}
\caption{\label{fig:chiralsusp}$\chi^{(\Delta \beta)} _{disc}$ as functions of temperature at different temperature gradients.
For each fixed $\Delta \beta$, the highest peaks of susceptibilities are marked with small circles.}
\end{center}
\end{figure}

\begin{figure}[htbp]
\begin{center}
\includegraphics[width=0.8\hsize]{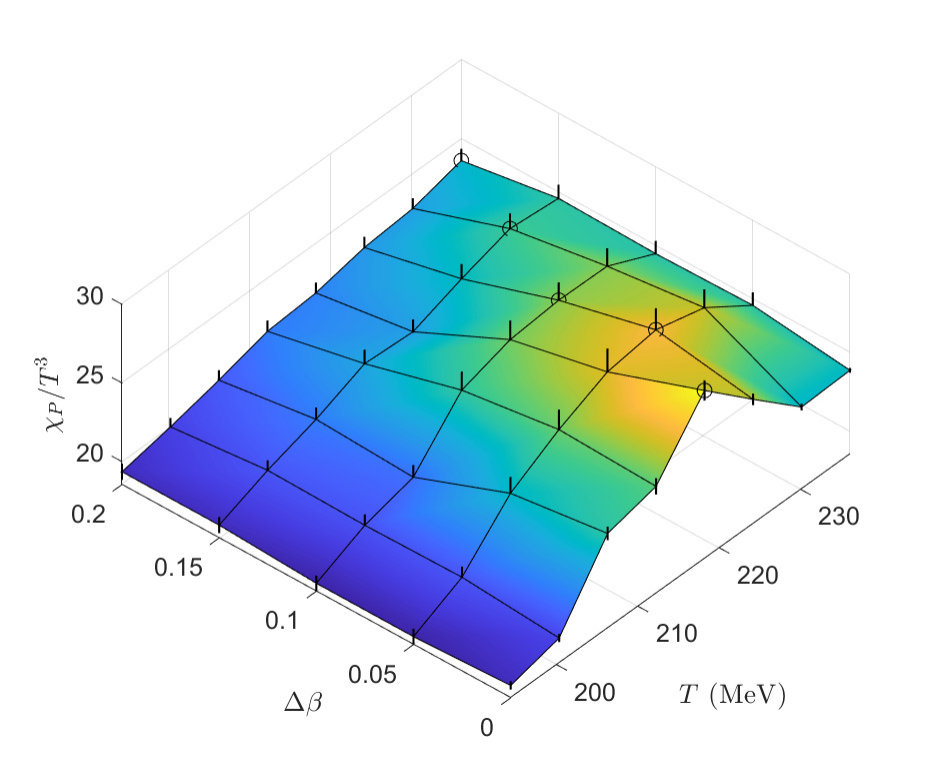}
\caption{\label{fig:polyasusp}Same as Fig.~\ref{fig:chiralsusp} but for $\chi^{(\Delta \beta)} _P$.}
\end{center}
\end{figure}

For the quenched approximation, the peaks of $\chi _{|P|}^{(\Delta \beta)}$ and $\chi _{\tilde{P}}^{(\Delta \beta)}$ are shown in Fig.~\ref{fig:quenchsusp}~(in this case the chiral condensation is not order parameter and therefore not investigated).
When dynamic fermions are turned on, the susceptibilities of the chiral condensation~($\chi^{(\Delta \beta)} _P$) and Polyakov loop~($\chi^{(\Delta \beta)} _{disc}$) as functions of temperature at different temperature gradients are shown in Figs.~\ref{fig:chiralsusp} and \ref{fig:polyasusp}.
It can be shown that, for $\Delta \beta =0$, the positions of the peaks of susceptibilities defined using $n_z=0,6$ are consistent with the susceptibilities of the whole lattice shown in Fig.~\ref{fig:suspuniform}.

For the quenched approximation, when $\Delta \beta=0.5$ the peak is pushed towards a higher temperature, which is consistent with the behavior that the temperature gradient will suppress the Polyakov loop in the deconfined phase, and is consistent with the model prediction~\cite{Zheng:2021pia}.

In Figs.~\ref{fig:chiralsusp} and \ref{fig:polyasusp}, for each fixed $\Delta \beta$, the highest peaks of susceptibilities are marked with small circles.
It can be seen that for the chiral condensation, the peak of the susceptibility moves to a lower temperature with the increase of the temperature gradient, meanwhile, for the Polyakov loop, the peak of the susceptibility moves to a higher temperature. 
In other words, the temperature gradient causes the pseudo-critical temperatures of the confine/deconfine transition and the chiral symmetry breaking transition to move away from each other.

As can be seen from the susceptibility of the Polyakov loop, the temperature gradient broadens and softens the peak of the susceptibility. 
Notice that for the $\Delta \beta=0$ case, Fig.~\ref{fig:polyasusp} shows the susceptibility also defined with $n_z=0,6$, so this broadening is not due to the smaller volume. 
Similar to the finite volume, a temperature gradient also pulls the system away from the singularity, which, together with the system's crossover feature, makes the order parameter more ill-defined. 
This may explain the fact that the shift of peaks of susceptibilities for chiral condensation are inconsistent with the change of chiral condensation. 

Temperature gradients that make the system more crossover can also lead to another consequence. 
Experiments have been devoted to finding the CEP in the $T-\mu$ plane, and at the same time, the temperature gradient is present in the fireballs in experiments. 
If the temperature gradient still makes the system more crossover at finite chemical potentials, then it can be expected that in experiments the CEP will be pushed in the direction of the first-order phase transition in the $T-\mu$ plane compared with theoretical predictions made for the case of uniform temperature, i.e., towards larger chemical potentials and lower temperatures.

It should be emphasized that the above phenomenon requires further efforts from the perspective of theoretical studies.

\subsection{\label{sec:3.4}Discussions of other effects}

Since our scheme for introducing temperature gradients is through the lattice spacing, it inevitably affects both the fermion mass, as well as the extents.
For the effect of the fermion mass, since we also studied the results under the quench approximation, i.e., when the fermion mass tends to infinity, it can be found that the temperature gradient still affects the system. 
Therefore, this subsection mainly considers the effect by the extents.
For this purpose, we consider another scheme that introduces a temperature gradient.

If one introduces anisotropic lattice spacing, i.e., $a_{x,y,z}=a, a_{\tau}=\xi(z)a$, the gauge action is than
\begin{equation}
\begin{split}
&S_G=\frac{\beta}{N_c}\sum _n\left\{\xi(z)\sum _{i>j}{\rm Retr}\left(1-\bar{U}_{ij}(n)\right)\right.\\
&\left.+\frac{1}{\xi(z)}\sum _{i}{\rm Retr}\left(1-\bar{U}_{i\tau}(n)\right)\right\},\\
\end{split}
\label{eq.3.newaction2}
\end{equation}
where the sum over the Latin index is the sum over $x,y,z$, and the definition of $\bar{U}_{\mu\nu}$ is the same as the one in Eq.~(\ref{eq.2.1}).
In the action defined in Eq.~(\ref{eq.3.newaction2}), one can adjust $a_{\tau}$ without changing the lattice spacing in the spatial dimension. 
If we assume that the effect on the matching of $a$ is negligible, then this is equivalent to changing the temperature while keeping extents unchanged.
In this way, one can avoid the effect of the gradient in extents by considering the effect of the temperature gradient alone.

In order to work with the periodic boundary conditions, we again use a trigonometric function for $\xi(z)$,
\begin{equation}
\begin{split}
&\xi(z)=\frac{1}{1+\Delta \xi \sin(\frac{2\pi n_z}{L_z})},\\
&\frac{\partial T}{\partial z}=\frac{2\pi \Delta \xi}{a^2 L_{\tau}L_z}\cos\left(\frac{2\pi n_z}{L_z}\right).\\
\end{split}
\label{eq.3.newaction3}
\end{equation}
At $n_z=0,6$, $\partial T/\partial z = \pi a^{-2} \Delta \xi /36$.
Similar to the previous sections, we define $P^{[\Delta \xi]}$, $\tilde{P}^{[\Delta \xi]}$ and $\chi _P^{[\Delta \xi]}$ as same as $P^{(\Delta \beta)}$, $\tilde{P}^{(\Delta \beta)}$ and $\chi _P^{(\Delta \beta)}$ but just replace $\Delta \beta$ with $\Delta \xi$ in the notation to indicate that they are measured with configurations obtained by action in Eq.~(\ref{eq.3.newaction2}).

\begin{figure}[htbp]
\begin{center}
\includegraphics[width=0.48\hsize]{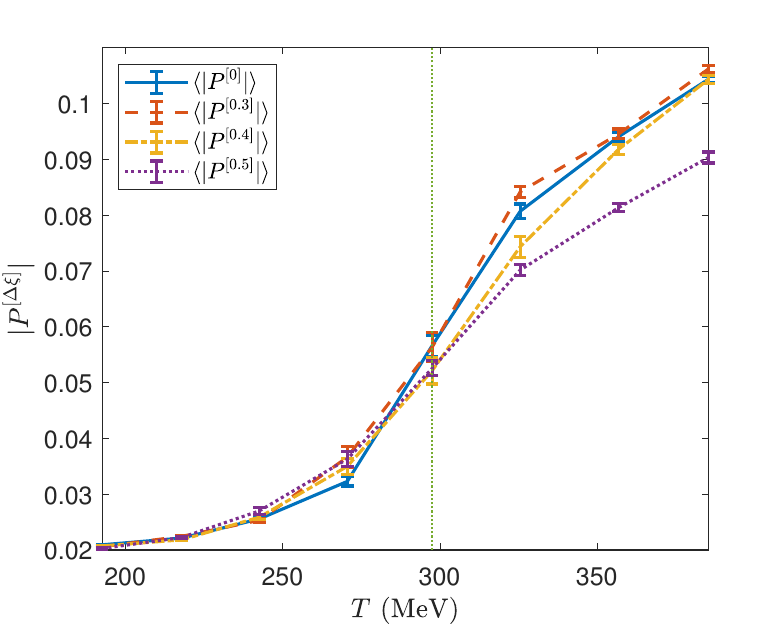}
\includegraphics[width=0.48\hsize]{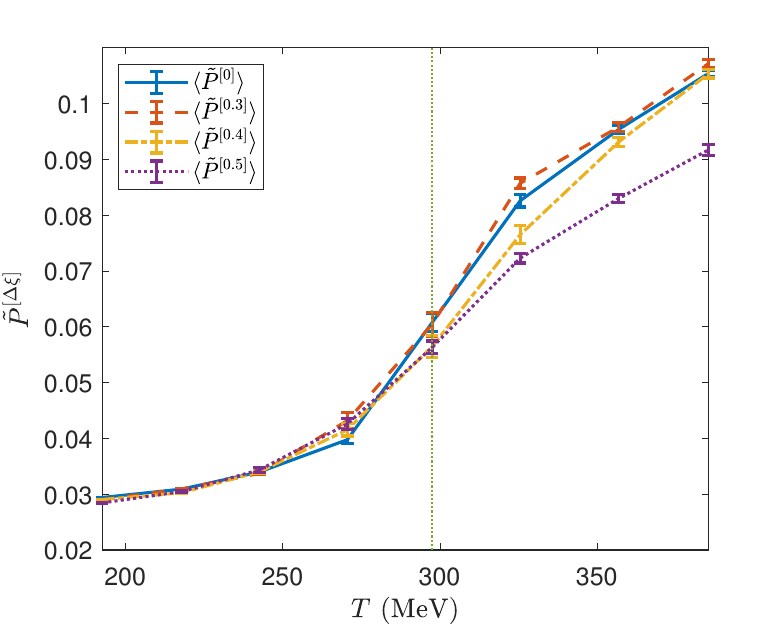}\\
\includegraphics[width=0.48\hsize]{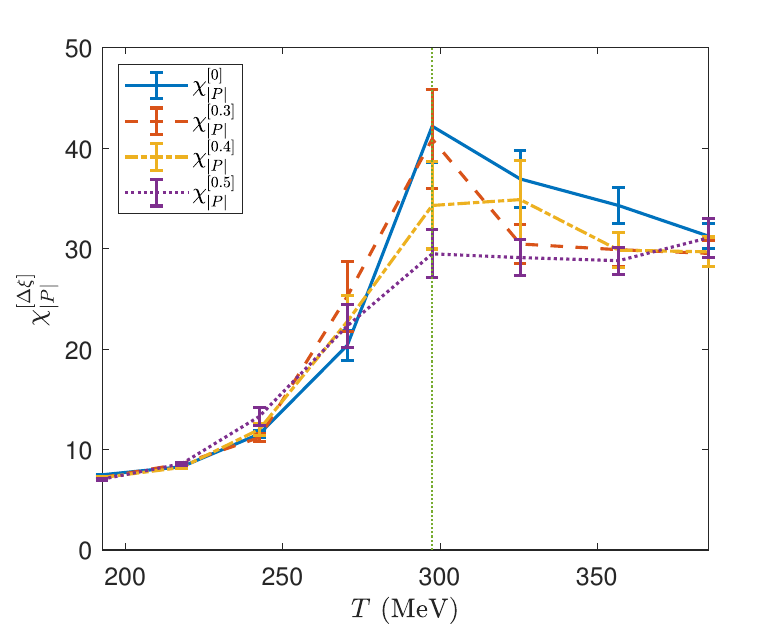}
\includegraphics[width=0.48\hsize]{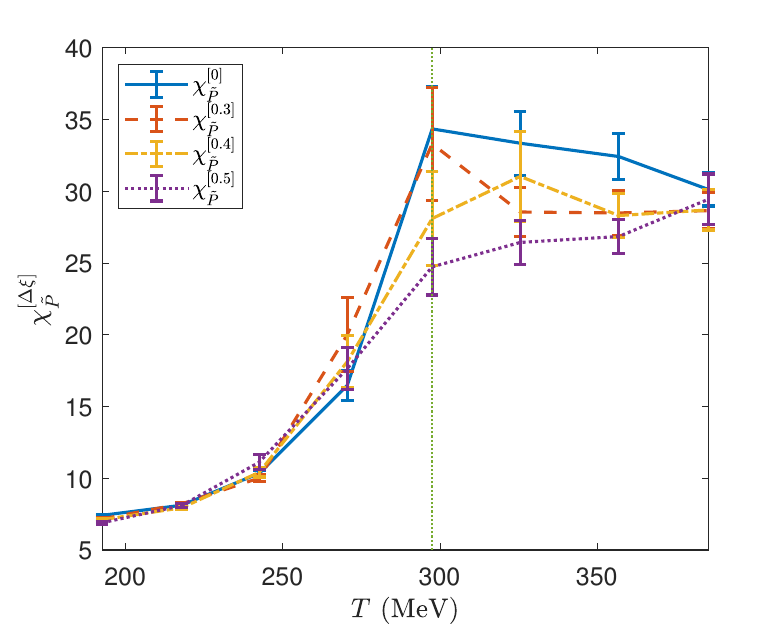}\\
\caption{\label{fig:quenchatgradient}$|P^{[\Delta \xi]}|$~(top-left panel), $\tilde{P}^{[\Delta \xi]}$~(top-right panel), $\chi^{[\Delta \xi]} _{|P|}$~(bottom-left panel) and $\chi^{[\Delta \xi]} _{\tilde{P}}$~(bottom-right panel) as functions of $T$ in quenched approximation.
The dotted lines show the critical temperature for the case of uniform temperature.}
\end{center}
\end{figure}
For each $\Delta \xi=0.3,0.4$ and $0.5$, $2000+(100+19900)\times 8$ TUs are simulated, where $20000$ TUs are simulated for each value of $\beta$ sequentially with $\beta$ growing, except for the first value of $\beta$ that $12000$ TUs are simulated. 
The last $19900$ configurations for each value of $\beta$ are measured. 
The results of $P^{[\Delta \xi]}$ and $\chi _P^{[\Delta \xi]}$ are shown in Fig.~\ref{fig:quenchatgradient}.
When $\Delta \xi=0.3$, the $\partial T/\partial z$ is from $(187\;{\rm MeV})^2$~(when $\beta=5.7$) to $(374\;{\rm MeV})^2$~(when $\beta=6.05$).
When $\Delta \xi=0.4$, the $\partial T/\partial z$ is from $(216\;{\rm MeV})^2$~(when $\beta=5.7$) to $(432\;{\rm MeV})^2$~(when $\beta=6.05$).
When $\Delta \xi=0.5$, the $\partial T/\partial z$ is from $(241\;{\rm MeV})^2$~(when $\beta=5.7$) to $(482\;{\rm MeV})^2$~(when $\beta=6.05$).

It can be seen from Fig.~\ref{fig:quenchatgradient} that with the gradual increase of the temperature gradient, a decrease in the value of Polyakov loop occurs in the deconfined phase. 
In addition, it can also be seen from the results of $\chi_P$ that the peak of $\chi _P$, which is the location of the critical temperature of the phase transition, is gradually pushed to a higher temperature region as the temperature gradient gradually increases.
Both behaviors are consistent with the previous results using $\beta(z)$, which shows that these behaviors are mainly due to the temperature gradient, but not to the effect of non-uniformity in volume.

\section{\label{sec:4}Summary}

In colliders, the fireball is at a location dependent temperature distribution instead of a uniform temperature.
In this work, the presence of a temperature gradient is studied using lattice QCD approach with location dependent $\beta$.

In the quenched approximation, the temperature gradient always catalyzes the chiral symmetry breaking.
Meanwhile, the Polyakov loop is suppressed by the temperature gradient, but this only happens in the deconfined phase, in the confined phase, the Polyakov loop is barely affected by the temperature gradient.
The temperature gradient will increase the critical temperature of confine/deconfine phase transition, which is consistent with the model prediction.

When dynamic fermions are turned on, the temperature gradient still catalyzes the chiral symmetry breaking.
However, the Polyakov loop is increased in the deconfined phase, while decreased in the confined phase, i.e., in the confined phase, the chiral condensation and the Polyakov loop are increased at the same time.
Not only that, but the pseudo-critical temperatures of the chiral symmetry breaking transition and the confine/deconfine transition move in the opposite directions.
Although the chiral symmetry breaking is catalyzed, the pseudo-critical temperature is lowered down.
The presence of temperature gradient broadens and softens the peaks of susceptibilities, if the phenomenon still holds at finite chemical potentials, then it is implied that the CEP in the $T-\mu$ plane will move in the direction of the first-order phase transition.
However, a fully understanding of these phenomena will require future theoretical studies.

\begin{acknowledgments}
This work was supported in part by the National Natural Science Foundation of China under Grants Nos. 11875157 and 12147214, and the Natural Science Foundation of the Liaoning Scientific Committee No.~LJKZ0978.
\end{acknowledgments}

\bibliography{tempgradient}% Produces the bibliography via BibTeX.

\end{document}